\documentclass[%
floatfix,
 reprint,
 amsmath,amssymb,
 aps,prl,showpacs
]{revtex4-1}

\usepackage{graphicx}
\usepackage{dcolumn}
\usepackage{bm}
\usepackage{todonotes}

\newcommand{\OpenFOAM}{OpenFOAM\textsuperscript{\tiny\textregistered}}

\begin{document}

\title{Controlling Elastic Turbulence}

\author{Reinier van Buel}
 \email{r.vanbuel@tu-berlin.de}
\author{Holger Stark}%
\affiliation{%
Technische Universit{\"a}t Berlin,
{Institute of Theoretical Physics,}
Hardenbergstrasse 36, 10623 Berlin, 
{Germany}
}%


\date{\today}

\begin{abstract}
We demonstrate through numerical solutions of the Oldroyd-B model in a two-dimensional Taylor-Couette geometry that
the onset of elastic turbulence in a viscoelastic fluid {can be controlled} by imposed shear-rate modulations.
{While for slow modulations elastic turbulence is still present, it vanishes for} fast modulations {and} a laminar response {with the Taylor-Couette base flow is recovered.}
We find that the transition from {the} laminar {to the} turbulent state is supercritical and occurs at a critical Deborah number.
{In the state diagram of}
both control parameters, {Weissenberg versus Deborah number, we identify the region of elastic turbulence.
We also quantify the transition by the flow resistance, which in the laminar regime we can describe within the (linear) Maxwell model.}


\end{abstract}


\maketitle


Controlling the flow pattern of viscoelastic fluids is extremely challenging due to their inherent non-linear 
properties and {their} strong response to shear deformations \cite{groisman2000,groisman2001stretching,squires2005microfluidics}.
Viscoelastic fluids, such as {polymer solutions,} exhibit transitions from laminar to {time-dependent} non-laminar flows, which is useful for heat and mass transport at the micron scale 
\cite{groisman2000,groisman2001stretching,groisman2004elastic, thomases2009transition,thomases2011stokesian,kumar1996chaotic,niederkorn1993mixing,arratia2006elastic}
{whereas} in Newtonian fluids {transport} {on such small scales} is dominated by diffusion. Turbulent viscoelastic {flow fields} show similar properties as {their counterparts in Newtonian fluids}
\cite{groisman2004elastic}.
Consequently, the state of the occurring flow pattern
is called \textit{elastic turbulence} \cite{groisman2000}. 
Since the discovery of this seminal effect at the beginning of the new millennium \cite{groisman2000},
research is ongoing 
{\cite{bodiguel2015flow,afik2017role,qin2017characterizing,belan2018boundary,varshney2019elastic,qin2019flow,steinbergscaling}}.
{The transition to elastic turbulence is accompanied by an enhanced drag resistance in flowing polymer solutions \cite{groisman2004elastic,groisman2000,groisman2001stretching,qin2019flow}.
In this letter we report on a method to reduce and ultimately prevent elastic turbulence by applying a time-modulated 
shear rate.}

Controlling flow patterns and fluid instabilities {in} Newtonian fluids 
has extensively been studied \cite{barenghi1989modulated,barenghi1991computations,kuhlmann1985model,lopez2002modulated,marques1997taylor,weisberg1997delaying,zeitz2015feedback}.
In con\-trast, the search for control strategies appropriate for viscoelastic fluids has so far been limited.
{For example, in}
Taylor-Couette geometries spatially modulated cylinders were used 
{to induce}
pattern formation \cite{anglade2005pattern}, 
while su\-per\-imposed axial flow delays the onset of the viscoelastic flow instability \cite{graham1998effect}.
In Ref.\ \cite{zhang2019modulation} different responses of a Poiseuille flow to periodically modulated driving were observed,
while disorder in microfluidic flows can inhibit elastic turbulence \cite{walkama2019disorder}. Finally, 
the authors of 
Ref.\ \cite{von2018time}  
were able to stabilize a micellar suspension against a shear-banding instability using 
time-delayed feedback\ \cite{pyragas1995control}.

In Newtonian fluids the transition to turbulence is solely driven by inertia and therefore characterized by the Reynolds 
number Re \cite{batchelor1977developments}. 
In the following we concentrate on small Reynolds numbers, where inertia can be neglected. Then the 
transition from steady laminar flow to elastic turbulence is determined by the Weissenberg number $\mathrm{Wi}$, the product of 
an intrinsic fluid relaxation time and the fluid deformation rate \cite{pakdel1996elastic,poole2012deborah}

Importantly, the critical Weissenberg number,
{at which this transition occurs,}
depends on the geometry and especially on the curvature of the flow streamlines \cite{pakdel1996elastic}. In {experimental} geometries with curved streamlines a purely elastic instability has been observed {in} Taylor-Couette flow \cite{larson1990purely, groisman2004elastic}, von K{\'a}rm{\'a}n swirling flow \cite{mckinley1991observations,byars1994spiral,groisman2000,groisman2004elastic,burghelea2007elastic}, serpentine channel or Dean flow \cite{groisman2004elastic,ducloue2019secondary}, cone-and-plate flow \cite{mckinley1991observations}, cross-channel flow \cite{arratia2006elastic,sousa2018purely}, and lid-driven cavity flows \cite{pakdel1996elastic}. 
{Viscoelastic fluids flowing through straight microchannels}
are linearly stable and non-linearly unstable \cite{morozov2005subcritical,pan2013nonlinear,morozov2015introduction}.
{Different numerical}
techniques 
{were}
employed 
{to solve}
constitutive equations modeling viscoelastic fluids 
{and thereby}
also 
revealed the purely elastic 
{instability} 
in similar geometries.
{Articles}
address sinusoidal forcing \cite{gupta2019effect,gutierrez2019proper,thomases2011stokesian}, Kolmogorov flow \cite{berti2008two,berti2010elastic}, 
{sudden-expansion flow}
\cite{poole2007plane}, as well as channels with cross-slot geometry \cite{poole2007purely} {or serpentines} 
\cite{poole2013viscoelastic}.
In our own simulations using the Oldroyd-B model in a 2D Taylor-Couette geometry we could confirm a supercritical transition 
above a critical Weissenberg number \cite{vanBuel2018elastic}.

Addressing time-dependent shear flows of viscoelastic fluids, which we will do in this article, needs another characteristic 
number. We will employ the Deborah number $\mathrm{De}$ and formulate it as the ratio of the intrinsic fluid 
relaxation time to the 
characteristic time of the deformation. It describes the degree of elastic response to 
{an external forcing applied}
over a given time frame 
\cite{poole2012deborah, dealy2010weissenberg}.

In this letter, we present first results on controlling the onset of turbulence in a viscoelastic fluid by means of modulating the 
applied shear rate. These results are important steps in applying control strategies to viscoelastic fluids.
We obtain numerical solutions of the Oldroyd-B model in a 2D Taylor-Couette geometry. 
{Since} our analysis {is restricted} to two spatial dimensions, we can investigate the simplest implementation of a Taylor-Couette flow. 
Although our {setting does not access the three dimensions} of {experimental flows,}
we can gain general insight into controlling viscoelastic fluids.
We have shown a transition to elastic turbulence at $\mathrm{Wi}=10$ in earlier work, where we applied a 
shear rate {constant in time}
in the same geometry \cite{vanBuel2018elastic}.
In this work, we {use} a {time-modulated} shear rate in the form of a square or sine wave.
We demonstrate how elastic turbulence is significantly reduced
{with increasing modulation frequency and ultimately vanishes at a critical Deborah number $\mathrm{De}_c$.
Here, the flow field assumes the radially symmetric base flow of the non-turbulent case.}

\begin{figure}
\centering
\includegraphics[width=0.45\textwidth]{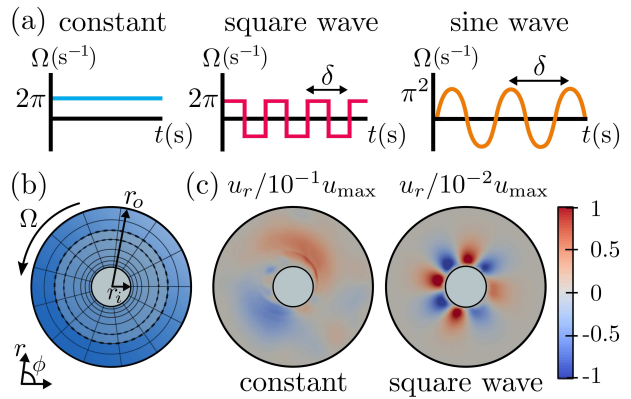}
  \caption{(a) %
{Angular}
velocity ${\Omega}$
{versus time}
{of}
the outer cylinder
{for different driving protocols.}
(b) Schematic of the 2D Taylor-Couette geometry, where $r_i$ is the radius of the inner cylinder, $r_o$ the radius of the outer cylinder and $\Omega$ the angular velocity
{of the outer cylinder.} 
(c) Color-coded radial component of the velocity field component
$u_r$ normalized by the maximum velocity $u_\text{max}$ for $\mathrm{Wi} = 21.4$. Left: at time
$t=225\,\mathrm{s}$, where $\Omega$ is constant. Right: at $t=375\,\mathrm{s}$ after the square-wave modulated driving with period $\delta = 12 \,\mathrm{s}$ has been switched on.
}
  \label{fig1}
\end{figure}

We examine
{the flow field $\bm{u}(\bm{r},t)$ of}
an incompressible viscoelastic fluid 
in a 2D Taylor-Couette geometry. The inner cylinder, at radius $r_i = 2.5 \mu m$, is fixed and the outer cylinder, at radius $r_o=4\; r_i$, rotates with 
{a}
periodically modulated angular velocity $\Omega$ with period $\delta$.
We distinguish between two different modulations: a square wave with amplitude 
\mbox{${\Omega_0 = 2\pi \, \mathrm{s}^{-1}}$} 
and a sine wave with amplitude 
\mbox{${\Omega_0 = \pi^2 \, \mathrm{s}^{-1}}$},
see Fig.~\ref{fig1}(a).
These amplitudes are chosen {such} that the Weissenberg numbers
{defined with $|\Omega|$ averaged} over one period are equal, 
$\mathrm{Wi}=1/\delta\int_{0}^{\delta} \lambda|{\Omega}|\mathrm{d}t = 2 \pi \lambda \, 
{\mathrm{s}^{-1}}$, 
with $\lambda$ the elastic relaxation time of the fluid. The Deborah number
{is determined by the rate of change in the shear flow and thus} is given by $\mathrm{De}= \lambda / \delta$.

To model the {viscoelastic} fluid, we use the Oldroyd-B model. {It uses the polymeric stress tensor}
\begin{align}
\label{eq:OLD} 
\bm{\tau} +\lambda \overset{\nabla}{\bm{\tau}} &= 
{
\eta_p \left[ \nabla \otimes \mathbf{u} + (\nabla \otimes \mathbf{u})^\mathrm{T} \right] \, ,
}
\end{align}
where $\eta_p$ {is} the polymeric shear viscosity and $\overset{\nabla}{\bm{\tau}}$ denotes the 
upper convective derivative of the stress tensor defined as
\begin{equation}
\overset{\nabla}{\bm{\tau}} = \frac{\partial  \bm{\tau}}{\partial t} + \left(\mathbf{u}\cdot \nabla\right)\bm{\tau} 
- (\nabla \otimes \mathbf{u})^\mathrm{T} \bm{\tau} - \bm{\tau}  ( \nabla \otimes \mathbf{u})  \, .
\end{equation}
The hydrodynamic continuity {equations for density and momentum read}
\begin{align}
\nabla \cdot \mathbf{u}  &= 0\, ,  \\
\label{eq:NS} 
\rho \Big(
\frac{\partial \mathbf{u}}{\partial t} + \left(\mathbf{u} \cdot \nabla\right) \mathbf{u}  \Big)
&= - \nabla p + \eta_s \nabla^2 \mathbf{u} +  
{\nabla \bm{\tau}} \, .
\end{align}
Here, $\rho$ is the density of the {incompressible}
solvent, $p$ the pressure, and $\eta_s$ the solvent shear viscosity. 
We set the solvent viscosity to {$\eta_s=10^{-3} \mathrm{kg/ms}$}
and the polymeric viscosity to $\eta_p=1.5\eta_s$.
{From the given para\-me\-ters we then calculate} a very small Reynolds number 
\mbox{$\mathrm{Re} = \rho \Omega r_o^2 / \eta_s\approx 10^{-4}$.}

We obtain numerical solutions of 
{the Eqs.}
(\ref{eq:OLD})-(\ref{eq:NS}) using the open-source program \OpenFOAM. We 
{show} a schematic of our computational mesh in Fig.\ \ref{fig1}(b).
{The} simulations 
{with a time-modulated driving of the outer cylinder}
are either started from rest (velocity and stress tensor fields {are zero)}
or from a turbulent state, which is obtained by applying {first} a constant rotation for 
{a time range of} $250\,\mathrm{s}$.
Details on the {algorithm,} 
mesh and mesh refinement can be found in Ref.\ \cite{vanBuel2018elastic}.
The flow {field} is characterized through its fluctuations by the secondary-flow strength
\begin{equation}
\sigma(t) \equiv \left. \sqrt{\left\langle [\mathbf{u}(\mathbf{r},t)-\mathbf{u}^0(\mathbf{r}, {t})]^2 \right\rangle_{r,{\phi}}} 
\right/u^0_{\mathrm{max}} \, ,
\label{Eq.standard}
\end{equation}
where we take the square root of the second moment of the flow field 
{relative to}
the base flow $\mathbf{u^0}(\mathbf{r,t})=u_\phi^0 = A r + B r^{-1}$, $A = \frac{r^2_o }{r^2_o-r^2_i}\Omega$, and 
$B = -\frac{r^2_i r^2_o}{r^2_o-r^2_i}\Omega$ \cite{larson1990purely}, 
where for $\Omega$ we take the periodically modulated angular velocity.
It solves the Oldroyd-B model in the laminar case without turbulence.
Here, $\langle ...\rangle_{r,\phi}$ denotes the spatial average over coordinates $r,\phi$.
For constant rotation of the outer cylinder and  $\mathrm{Wi}>\mathrm{Wi}_c=10$, the secondary-flow strength is increasingly irregular with increasing $\mathrm{Wi}$ \cite{vanBuel2018elastic}.
The radial symmetry of the flow is broken and a radial velocity component $u_r(\phi)$ emerges [see Fig. \ref{fig1} (c), left].
Furthermore, the transition from a laminar to a turbulent flow state is supercritical \cite{vanBuel2018elastic}.

\begin{figure}
\centering
\includegraphics[width=0.42\textwidth]{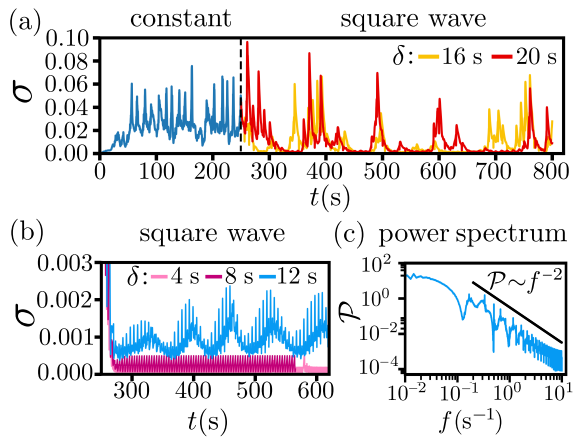}
  \caption{
  (a) and (b) Secondary-flow strength $\sigma$ as a function of time $t$. The outer cylinder rotates with a constant angular velocity $\Omega=2\pi \,\mathrm{s}^{-1}$ for the first $250 \,\mathrm{s}$. 
{Then the}  
modulated square-wave 
{driving}
with period $\delta$ and amplitude 
${\Omega_0}
= 2 \pi \, \mathrm{s^{-1}}$
{is switched on. The Weissenberg number is $\mathrm{Wi}=21.4$ resulting from the}
characteristic relaxation time $\lambda=3.4 \mathrm{s}$.
(c) Temporal power spectrum of secondary-flow strength, $\mathcal{P}=|\mathcal{F}(\sigma)|^2$, for $\delta=12\; \mathrm{s}$.
The observed spectrum scales as $\mathcal{P}\sim f^{-2}$.
}
  \label{fig2}
\end{figure}

The secondary-flow strength can be significantly lowered by applying a 
square-wave {driving to the outer cylinder as a comparison}
to the case of constant rotational velocity {shows} in Fig.~\ref{fig2} [see also Fig. 1(d), right].
For Weissenberg number $\mathrm{Wi}=21.4$ we present the secondary-flow strength for driving periods chosen from the range $4 \;\mathrm{s} \leq \delta \leq 20 \mathrm{s}$
{and observe} that {it} decreases with $\delta$. For lower frequencies ($\delta = 16\;\mathrm{s}$ or $ \delta = 20 \;\mathrm{s}$) $\sigma$ {exhibits} irregular peaks in time, which have magnitudes {comparable} to the case of constant rotation.
However, in between the irregular peaks the magnitude of $\sigma$ is much smaller and fluctuations in the flow are suppressed.
For high frequencies {$\sigma$ strongly tends to zero. It shows oscillations with a period equal to the driving period $\delta$}
{[see Fig.~\ref{fig2}(b)], before the flow ultimately becomes laminar.} Moreover, at $\delta=12\, \mathrm{s}$ the amplitude of the fast oscillations of the secondary-flow strength {seems to be modulated periodically. However,} {the power spectrum of $\sigma(t)$}
does not reveal such a regular modulation but shows an exponential decay of the correlations in $\sigma(t)$ with a characteristic time scale of the order of $\delta$. 
Superimposed on the exponential decay are fast modulations, the frequencies of which are multiples of $\delta^{-1}$, see Fig.~\ref{fig2}(c).

\begin{figure}
\centering
\includegraphics[width=0.38\textwidth]{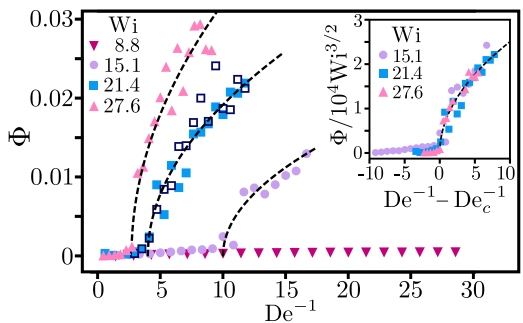}
  \caption{
Order parameter $\Phi$ as a function of the inverse Deborah number $\mathrm{De}^{-1}=\delta/\lambda$ in the case of square wave modulations for four Weisenberg numbers.
The time average of the secondary-flow strength is taken over at least $500\,\mathrm{s}$ in the turbulent regime; after the flow has been driven for $250\,\mathrm{s}$ with a constant velocity. 
{Open blue squares: the modulated driving starts from the beginning.}
The {dashed} lines are square root fits {to} $\Phi \sim \sqrt{\mathrm{De}^{-1} - \mathrm{De}_c^{-1}}$.
{Inset: the rescaled} data collapse onto a single master curve.
  }
  \label{fig3}
\end{figure}
\begin{figure}
\centering
\includegraphics[width=0.38\textwidth]{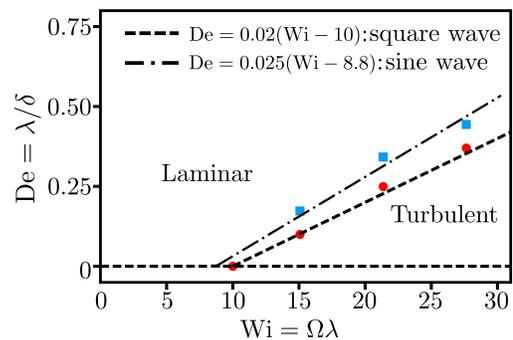} 
  \caption{
State diagram {showing} the state of the viscoelastic fluid as a function of the Weissenberg number {$\mathrm{Wi}=\Omega\lambda$} and Deborah number {$\mathrm{De}=\lambda/\delta$}.
The transition between the laminar and turbulent states is demarcated by the dashed line 
$\mathrm{De} = 0.02 ( \mathrm{Wi} - 10)$ 
for square wave modulation and by the black dashed-dotted line 
$\mathrm{De} = 0.025 ( \mathrm{Wi} - 8.8)$
for sine wave modulation {\cite{footnote}.}
The observed critical values are indicated with red circles and blue squares, respectively.
}
  \label{fig6}
\end{figure}

In Fig.\ \ref{fig3} we plot the order parameter, defined as the time average of the secondary-flow strength, $\Phi = \overline{\sigma}$,
versus the inverse Deborah number for different Weissenberg numbers under square wave driving. It sharply increases above a critical {value} $\mathrm{De}_c^{-1}$, which depends on $\mathrm{Wi}$. The transition scales as $(\mathrm{De}^{-1}- \mathrm{De}_c^{-1})^{1/2}$ implying that 
it is supercritical. This result is further tested by applying 
the modulated driving directly to the rest state (open square symbols for $\mathrm{Wi} = 21.4$ in Fig.\ \ref{fig3}).
The different initial conditions do not lead to different values of $\Phi$, as is expected for a supercritical transition.
We also checked that the supercritical transition occurs for sinusoidal driving.
The order parameter $\Phi$ displays qualitatively similar behavior. 
However, the critical values $\mathrm{De}^{-1}_c$ are smaller compared to the square wave driving for the same $\mathrm{Wi}$ and $\Phi$ quickly reaches a maximum value.
The results are presented in the supplemental material.
Another striking feature is that the order parameter displays a universal behavior
around the transition. 
Indeed, as the inset of Fig.~\ref{fig3} demonstrates, all curves for different $\mathrm{Wi}$ fall on a single master curve when we normalize $\Phi$ by $\mathrm{Wi}^{3/2}$ and plot them versus $\mathrm{De}^{-1} - \mathrm{De}_c^{-1}$.
To illustrate how the transition towards elastic turbulence depends on both dimensionless numbers ($\mathrm{De}$, $\mathrm{Wi}$), 
we have plotted the state diagram in Fig.~\ref{fig6} {for both modulation types}. It clearly demonstrates how the region of elastic turbulence is reduced upon increasing the Deborah number,
{meaning when the frequency of the modulated driving is increased.}

\begin{figure}
\centering
\includegraphics[width=0.38\textwidth]{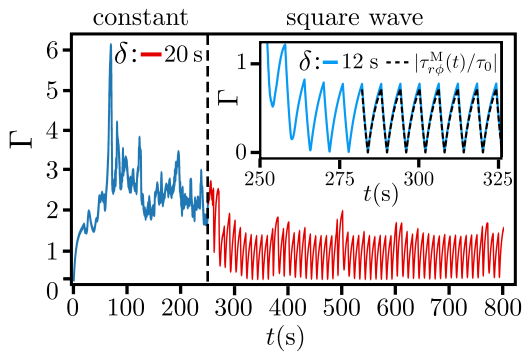} 
  \caption{The flow resistance
 at the outer cylinder,
$\Gamma = \left\langle \left | \tau_{r\phi}(r_o) /\tau_{r\phi}^0(r_o)  \right | \right \rangle_{\phi}$,
plotted versus $t$ at $\mathrm{Wi}=21.4$. The square wave driving
with period $\delta = 20 \,\mathrm{s}$ starts at $t=250 \, \mathrm{s}$.
Inset: $\delta = 12 \,\mathrm{s}$. The dashed line indicates the analytic result $|\tau^{\mathrm{M}}_{r\phi}(t)/\tau_0| $, 
where $\tau^{\mathrm{M}}_{r\phi}(t)$ is given by Eq. (\ref{eq.gammaM}).
}
  \label{fig4}
\end{figure}

\begin{figure}
\centering
\includegraphics[width=0.38\textwidth]{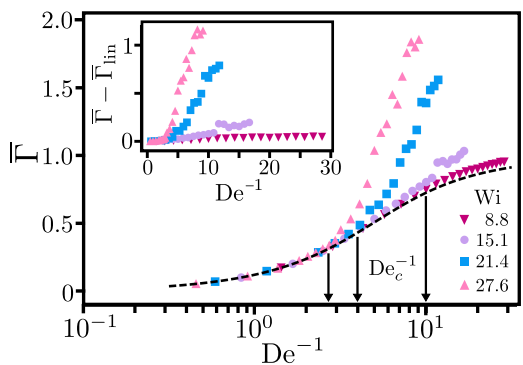}
  \caption{Time-averaged flow resistance or polymeric shear stress at the outer cylinder, $\overline{\Gamma}$, plotted versus the inverse Deborah number $\mathrm{De}^{-1}=\delta/\lambda$, for the same parameters as in Fig.~\ref{fig3}.
  The dashed line indicates $\overline{\Gamma}_{\mathrm{lin}}$ for the linear stress response of a Maxwell fluid. The arrows indicate $\mathrm{De}^{-1}_c$ from right to left for $\mathrm{Wi} = 15.1$, $21.4$, and $27.6$. Inset: $\overline{\Gamma} - \overline{\Gamma}_{\mathrm{lin}}$ versus  $\mathrm{De}^{-1}$.
}
  \label{fig5}
\end{figure}

The elastic nature of the transition to elastic turbulence can also be monitored by the polymeric shear stress $\tau_{r\phi}$,
which, when calculated at the outer cylinder, serves as an experimentally accessible measure for the flow resistance. For the
steady azimuthal base flow the shear stress component becomes $\tau_{r\phi}^0 = -2 \eta_p B r^{-2}$ \cite{larson1990purely}.
Note that it does not depend on $\lambda$ or $\mathrm{Wi}$. Thus for the azimuthal flow the non-linear terms in the constitutive relation (\ref{eq:OLD}) of the polymeric stress tensor are not relevant. Now, we introduce the flow resistance using the shear stress at the
outer cylinder ($r=r_o$),
\begin{equation}
\Gamma \equiv \left\langle \left |    \tau_{r\phi}(r_o) /\tau_{r\phi}^0(r_o)  \right | \right \rangle_{\phi}  \, .
\label{Eq.stress}
\end{equation}
For the steady laminar base flow, $\Gamma=1$, as defined. Under constant driving for $\mathrm{Wi}>\mathrm{Wi}_c$,
elastic turbulence with a radial secondary flow develops\ \cite{vanBuel2018elastic}. Through the non-linear terms in 
Eq.\ (\ref{eq:OLD}) all polymeric stress components couple to each other and one has $\Gamma > 1$. This is illustrated in
Fig.\ \ref{fig4} until time $t=250 \mathrm{s}$. Then the square-wave driving is switched on. For the period $\delta = 20 \mathrm{s}$
corresponding to $\mathrm{De}^{-1}=5.9>\mathrm{De}_c^{-1}$, $\Gamma$ is reduced but still reaches values above one and its
time evolution is still irregular, as expected for the turbulent state. In contrast, decreasing $\delta$ further to $12 \mathrm{s}$ (or $\mathrm{De}^{-1}=3.5 < \mathrm{De}_c^{-1}$), $\Gamma$ becomes regular and can 
be fit well by the linear version of the Oldroyd-B model (see next paragraph). Thus the laminar state of the base flow is recovered 
as also indicated in Fig.\ \ref{fig3}.

We can now add some understanding for the control of elastic turbulence under modulated driving. For sufficiently large 
period $\delta$ or $\mathrm{De}^{-1}$, the polymer elastic stress has sufficient time to build up,
generate the necessary ``hoop stress'' \cite{pakdel1996elastic,groisman2004elastic},
and thereby ultimately induce elastic turbulence.
However, this is no longer possible for fast switching between negative (clockwise) and positive 
(counter clockwise) driving. The dissolved polymers can only react with small elongations similar to the fast driving of a harmonic oscillator and the generated stress is not sufficient for elastic turbulence to occur.

{To quantify this argument further, we compare the polymeric shear stress 
to the shear stress of the linear Oldroyd-B or Maxwell model. Note again, the linear model applies to $\tau_{r\phi}$ when calculated
for the azimuthal base flow. Thus we solve
$\lambda \dot{\tau}_{r \phi}^{\mathrm{M}} + \tau_{r \phi}^{\mathrm{M}} = \eta_p \dot{\gamma}(t)$ for a shear rate switching 
periodically between $\pm \dot{\gamma}_0$. Using the formalism of Green's function as detailed in the supplemental material,
we arrive at
\begin{equation}
\frac{\tau_{r\phi}^{\mathrm{M}}(t)}{\tau_0} = \pm 
\bigg( 1 - 2 \frac{\mathrm{e}^{-(t-t_s)/\lambda}}{1+\mathrm{e}^{-\mathrm{De}^{-1}/2}} \bigg) \, , \enspace  t_s \le t \le t_s + \delta/2
\label{eq.gammaM}
\end{equation}
where $\pm$ means that the applied shear rate has switched to $\pm \dot{\gamma}_0$ at time $t_s$ and 
$\tau_0 = \eta_p  \dot{\gamma}_0$ is the shear stress of the base flow.}
Now we consider the time-averaged flow resistance $ \overline{\Gamma}$, with $\Gamma$ defined in Eq.\ (\ref{Eq.stress}).
The corresponding quantity for the Maxwell fluid can be calculated using the periodic solution from Eq.\ (\ref{eq.gammaM}):
$\overline{\Gamma}_\mathrm{lin} =  \frac{2}{\delta} \int_{t_s}^{t_s+\delta/2} |\tau^{\mathrm{M}}_{r\phi}(t)/\tau_0|  \mathrm{d}t 
= 1 + 4\, \mathrm{De} \ln\left( [1+\exp(- \mathrm{De}^{-1}/2)]/2\right)$.

Figure \ref{fig5} {plots} $\overline{\Gamma}$ versus $\mathrm{De}^{-1}$ for different $\mathrm{Wi}$ together with the analytic 
result $\overline{\Gamma}_{\mathrm{lin}}$ for the Maxwell fluid as the dashed line. Below the critical $\mathrm{De}_c^{-1}$ the 
data fall on $\overline{\Gamma}_{\mathrm{lin}}$ clearly indicating that for sufficiently fast modulation the linear response of the 
laminar base flow is recovered. Above $\mathrm{De}_c^{-1}$ we observe a significant increase in the stress response 
$\overline{\Gamma}$ due to the turbulent flow. In contrast, for $\mathrm{Wi}=8.8 < \mathrm{Wi}_c$ the flow remains laminar 
and  $\overline{\Gamma}\approx \overline{\Gamma}_\mathrm{lin}$.
The slightly larger values of the numerical stress data result from the non-linear terms in the constitutive relation of the stress tensor so that the simulated flow field deviates from the base flow.
Again, similar behavior is observed in the case of sinusoidal driving (presented in the supplemental material).
However, $\overline{\Gamma}$ increases more rapidly with $\mathrm{De}^{-1}$ than for square-wave driving.

In summary, by modulating the shear rate of a Taylor-Couette flow 
{using}
a square- or sine-wave 
{driving of}
the outer cylinder, we
{are able to control the onset of elastic turbulence. While at small frequencies (small Deborah numbers) 
irregular flow patterns are still observed in our simulations of the Oldroyd-B model, we recover the regular base flow at large
frequencies beyond a critical De. Here the (linear) Maxwell model accurately describes the rheological response of our system.}
{We hope our work paths the way for further investigations on the control of elastic turbulence and, in particular, initiates
appropriate experiments.}

\acknowledgments
{We thank A. Lindner and M. Wilczek
for 
stimulating discussions and acknowledge support from the Deutsche Forschungsgemeinschaft in the framework of 
the Collaborative Research  Center SFB 910.}

\bibliography{literature}

\begin{thebibliography}{51}
\bibitem [{foo()}]{footnote}%
  \BibitemOpen
  \href@noop {} {\enquote {\bibinfo {title} {Note, for the sine wave modulation
  the value $\mathrm{Wi}_c = 8.8$ for $\delta \to \infty$ is an approximation.
  \mbox{We} did not see any elastic turbulence at this value.}}\ }\BibitemShut
  {NoStop}%
\end{thebibliography}
\bibliographystyle{apsrev4-1}

\end{document}